\documentclass[11pt,showpacs,preprintnumbers,amsmath,amssymb]{revtex4}
\usepackage{amsfonts}
\usepackage{latexsym}
\begin{document}

\title{RKKY interaction in framework of $T=0$ Green function method}
\author{Todor M. Mishonov}
\author{Tihomir I. Valchev}
\author{Liliya A. Atanasova}
\author{Peter A. Ivanov}
\affiliation{Department of Theoretical Physics, Faculty of
Physics, Sofia University St.\,Kliment Ohridski, Bulgaria.}

\begin{abstract}
A methodical derivation of RKKY interaction in framework of $T=0$
Green function method is given in great detail. The article is
complimentary to standard textbooks on the physics of magnetism
and condensed matter physics. It is shown that the methods of
statistical mechanics gives a standard and probably simplest
derivation of the exchange interaction. A parallel with theory of
plasma waves demonstrates the relation between the Fourier
transformation of polarization operator of degenerate electron gas
at zero frequency and the space dependence of the indirect
electron exchange due to itinerant electrons.
\end{abstract}


\maketitle

The RKKY interaction is a basic ingredient in the physics of
magnetism, see for example the textbook by Mattis \cite{Mattis}.
The purpose of the present paper is purely methodical: to derive
RKKY result using the Green function technique for $T=0$. The
diagram formalism is described in the monograph by Lifshitz and
Pitaevski \cite{Lifshitz9} or the monograph by Abrikosov, Gor'kov
and Dzyaloshinski \cite{Abrikosov}. The s-d exchange interaction
\begin{equation}
 \hat H_{\mathrm{s-d}}= -J_{\mathrm{s-d}}
 \left( \hat{\mathbf{s}}(\mathbf{R_1})\cdot \mathbf{S_1}
      + \hat{\mathbf{s}}(\mathbf{R_2})\cdot \mathbf{S_2} \right)
\end{equation}
is treated as a small perturbation. Where $\mathbf{R_1}$ and
$\mathbf{R_2}$ are the space-vectors of the impurities,
$\mathbf{S_1}$ and $\mathbf{S_2}$ are their spins, and
$\hat{\mathbf{s}}(\mathbf{R_1})$ and
$\hat{\mathbf{s}}(\mathbf{R_2})$ are electron spin operators in
the corresponding space points. Using the simple matrix equation
\begin{align}
\mathrm{Tr}(\mathbf{\sigma}\cdot\mathbf{S_1}\mathbf{\sigma}\cdot\mathbf{S_2})=
  2\mathbf{S_1}\cdot\mathbf{S_2}
\end{align}
and the general rules for single loop approximation we obtain the
space dependence of the indirect exchange
\begin{equation}
\hat H_{\mathrm{RKKY}}= -J(\mathbf{R}) \mathbf{S_1}\cdot\mathbf{S_2},
\qquad \mathbf{R}=\mathbf{R}_2-\mathbf{R}_1,
\quad R= |\mathbf{R}|,
\end{equation}
where
\begin{align}
J(R)=\!\left(\frac{J_{\mathrm{s-d}}}{2}\right)^2
\!\left(\frac{a_0}{2\pi}\right)^6\mspace{-10mu}
\int\limits_{E(k)<\mu}\mspace{-17mu}d^{\,3}k\mspace{-7mu}
\int\limits_{E(k')>\mu}\mspace{-17mu}d^{\,3}k'
\frac{2\cos((\mathbf{k}-\mathbf{k'})\cdot\mathbf{R})}
{\frac{\hbar^2}{2m^*}(\mathbf{k'}^2-\mathbf{k}^2+i2m^*/\hbar\tau)},
\end{align}
where $\mu=\frac{k_F^2\hbar^2}{2m^*}$ is the Fermi energy and
$\hbar k_F$ is the Fermi momentum. This integral is actually the
Fourier transformation of the polarization operator of 3
dimensional electron gas for zero frequency. For a complementary
Green function derivation of the polarizability of degenerate
electron gas see the recent preprint \cite{plasmons}.

For simplification of further notation we introduce the constant
which parameterize the electron
scattering rate
$q^2=2m^*/\hbar\tau$.
Using the elementary relation
$$\cos((\mathbf{k}-\mathbf{k'})\cdot\mathbf{R})=
\cos(\mathbf{k}\cdot\mathbf{R})
\cos(\mathbf{k'}\cdot\mathbf{R})
+\sin(\mathbf{k}\cdot\mathbf{R})\sin(\mathbf{k'}\cdot
\mathbf{R})
$$
the integral above can be presented as
\begin{align}
 \int\limits_{E(k)<\mu}\mspace{-17mu}d^{\,3}&k\mspace{-7mu}
  \int\limits_{E(k')>\mu}\mspace{-17mu}d^{\,3}k'
  \frac{\cos((\mathbf{k}-\mathbf{k'})\cdot\mathbf{R})}
  {\mathbf{k'}^2-\mathbf{k}^2+iq^2}\,=I_1+I_2.
\end{align}
In order to solve these integrals we introduce spherical coordinates in
momentum spaces
$$
\mathbf{k}=k(\cos\varphi\sin\theta,\sin\varphi\sin\theta,\cos\theta),
\qquad \mathbf{R}=R (0,0,1),
$$
$$d^3k=k^2\sin\theta\,dp\,d\theta\,d\varphi,\qquad
\mathbf{k}\cdot\mathbf{R}=k\,R\cos\theta.
$$
One can easily check that due to the symmetry of integrant the
integration on $k'$ can be expanded to $k'<k_F.$
Then
\begin{align}
 I_1=&\int_0^{2\pi}\int_0^\pi\int_0^{k_F}d\varphi d\theta dkk^2\sin\theta
   \int_0^{2\pi}\int_0^\pi \int_0^{\infty}
   d\varphi'd\theta' dk'k^{\prime 2}\sin\theta'\\[7pt]
   &\times\frac{\sin(\mathbf{k}\cdot\mathbf{R})
   \sin(\mathbf{k'}\cdot\mathbf{R})}
   {\mathbf{k'}^2-\mathbf{k}^2+iq^2}\\[7pt]
  =&(2\pi)^2\int_0^\pi\int_0^{k_F}d\theta dk
   k^2\sin\theta\sin(k\,R\cos\theta)\\[7pt]
&\times\int_0^{\infty}dk'\frac{k^{\prime 2}}{\mathbf{k'}^2-\mathbf{k}^2+iq^2}
   \int_0^\pi d\theta'\sin\theta'\sin(k'R\cos\theta).
\end{align}
And using
\begin{align}
  \int_0^\pi d\theta\sin\theta\sin(kR\cos\theta)=
  \frac{1}{k\,R}\cos(k\,R\cos\theta)\bigg|_\pi^0=0,
\end{align}
we find that
\begin{align}
 I_1=0.
\end{align}
Analogously
\begin{align}
 I_2=&\int_0^{2\pi}\int_0^\pi\int_0^{k_F}d\varphi d\theta dkk^2\sin\theta
   \int_0^{2\pi}\int_0^\pi \int_0^{\infty}
   d\varphi'd\theta' dk'k^{\prime 2}\sin\theta'\\[7pt]
   &\times\frac{\cos(\mathbf{k}\cdot\mathbf{R})
   \cos(\mathbf{k'}\cdot\mathbf{R})}
   {\mathbf{k'}^2-\mathbf{k}^2+iq^2}\\[7pt]
  =&(2\pi)^2\int_0^\pi\int_0^{k_F}d\theta dk
   k^2\sin\theta\cos(k\,R\cos\theta)\\[7pt]
 &\times\int_0^{\infty}dk'\frac{k\prime^2}
  {\mathbf{k'}^2-\mathbf{k}^2+iq^2}
   \int_0^\pi d\theta'\sin\theta'\cos(k'R\cos\theta)
\end{align}
and
\begin{align}
 \int_0^\pi d\theta\sin\theta\cos(k\,R\cos\theta)=
  \frac{1}{k\,R}\sin(k\,R\cos\theta)\bigg|_\pi^0=
  \frac{2\sin(k\,R)}{k\,R}
\end{align}
we get
\begin{align}
 I_2=\left(\frac{4\pi}{R}\right)^2\int_0^{k_F}dk\;k\sin(k\,R)
  \int_0^{\infty}dk'\frac{k'\sin(k'R)}
   {\mathbf{k'}^2-\mathbf{k}^2+iq^2}.
\end{align}
In order to calculate the last integral we will apply the residuum
theorem for the function
\begin{align}
f(z)=\frac{z\exp(izR)}{z^2-k^2+iq^2},
\end{align}
i.e.
\begin{align}
\oint_\Gamma f(z)dz=2\pi i\sum_n\mathrm{Res}f(z_n), \qquad 1/f(z_n)=0,
\end{align}
we suppose that function $f$ has simple poles.
The contour contains the real axe  and infinite arch in upper semiplane;
$k'=\Re z.$
The arch gives zero part when it radius goes to infinity and the real axe
integral just presented by the residuum
\begin{align}
\int_{-\infty}^{\infty}f(z)dz=2\pi i\sum_n\mathrm{Res}f(z_n).
\end{align}
The poles of $f(z)$ obeys the equation
$$z^2=k^2-iq^2,$$
which has the solutions
\begin{align}
z_{1,\,2}=\sqrt[4]{k^4+q^4}\;e^{i(\alpha+n\pi)},\qquad
2\alpha=-\arctan\frac{q^2}{k^2}\qquad n=0,\,1.
\end{align}
Only $z=z_2$ belongs to upper semiplane and its residuum is
\begin{align}
\mathrm{Res}f(z=z_2)=\lim_{z\rightarrow z_2}(z-z_2)f(z)=
\frac{z_2\,e^{iz_2R}}{z_2-z_1}=\frac{e^{iz_2R}}{2}.
\end{align}
In weak scattering limit
$$q \ll k_{F}, \qquad |2\alpha|=\frac{q^2}{k^2}\ll 1,
\qquad z_2\approx -k(1+i\frac{q^2}{2k^2})
$$
and
\begin{align}
\mathrm{Res}f(z_2)\approx\frac{1}{2}e^{-ikR}e^{-q^2 R/2k}.
\end{align}
Then
\begin{align}
 \int_{-\infty}^{\infty}f(z)dz=i\pi e^{-ikR}e^{-q^2 R/2k}
\end{align}
and
\begin{align}
 \int_{0}^{\infty}dk'\frac{k'\sin(k'R)}
 {\mathbf{k'}^2-\mathbf{k}^2+iq^2}=
 \frac{1}{2}\Im\left(\int_{-\infty}^{\infty}f(z)dz\right)=
 \frac{\pi}{2}\cos(k\,R)e^{-q^2 R/2k}.
\end{align}
In the same weak scattering limit
$$\frac{q^2}{2k} \approx \frac{q^2}{2k_F}=\frac{1}{\lambda}$$
where $\lambda= \hbar k_F\tau/m^*$ is the electron mean free path,
we have
\begin{align}
 I_2=\left(\frac{4\pi}{R}\right)^2\frac{\pi}{2}\,e^{-R/\lambda}
 \int_0^{k_F}dk\;k\sin(k\,R)\cos(k\,R).
\end{align}
The elementary integration gives
\begin{align}
 &\int_0^{k_F}dk\;k\sin(k\,R)\cos(k\,R)=
 \frac{1}{2}\int_0^{k_F}dk\;k\sin(2k\,R)\\
 &=-\frac{k_F\cos(2k_FR)}{4R}+
 \frac{1}{4R}\int_0^{k_F}dk\cos(2k\,R)\\
 &=\frac{\sin(2k_FR)}{8R^2}-
 \frac{k_F\cos(2k_FR)}{4R}.
\end{align}
And for the second integral we obtain
\begin{align}
 I_2=\left(\frac{4\pi}{R}\right)^2\frac{\pi}{2}\,e^{-R/\lambda}
 \frac{\sin(2k_FR)-2k_FR\cos(2k_FR)}{8R^2}.
\end{align}
Finally for the exchange integral we arrive at
\begin{align}
 J(R)=(J_{\mathrm{s-d}})^2&
  \left(\frac{a_0}{2\pi}\right)^6\frac{\pi m^*}{2\hbar^2}
  \left(\frac{4\pi}{R}\right)^2e^{-R/\lambda}\\[7pt]
 &\times\frac{\sin(2k_FR)-2k_FR
  \cos(2k_FR)}{8R^2}\\[7pt]
 =(J_{\mathrm{s-d}})^2&\left(\frac{a_0k_F}{2}\right)^6
  \frac{8}{\pi^3\mu}\;e^{-R/\lambda}\\[7pt]
 &\times\frac{\sin(2k_FR)-2k_FR\cos(2k_FR)}{(2k_FR)^4}
\end{align}

Comparing with other explanations we come to the conclusion that
the method of $T=0$ Green functions give a standard and probably
simplest explanation of the RKKY interaction in metallic alloys. A
derivation directly based on the methods of statistical mechanics.
If we used temperature Green functions applied to
$T\rightarrow 0$ we would arrive at the the same residuum.

\end{document}